\documentclass[aps,prd,twocolumn,groupedaddress,showpacs,showkeys]{revtex4}

\usepackage{graphicx}
\usepackage{graphicx}
\usepackage{amssymb}
\usepackage{amsmath}
\usepackage{color}
\newcommand{\be}{\begin{equation}}
\newcommand{\ee}{\end{equation}}
 
 \definecolor{BrickRed}{cmyk}{0,0.89,0.94,0.28}
\definecolor{MidnightBlue}{cmyk}{0.98,0.13,0,0.43}
\definecolor{DarkGreen}{rgb}{0,0.7,0.1}

\begin{document}

\title{Casimir--Polder force between anisotropic nanoparticles and gently curved surfaces}

\date{\today}

\author{Giuseppe Bimonte}
\affiliation{Dipartimento di Fisica, Universit\`{a} di
Napoli Federico II, Complesso Universitario
di Monte S. Angelo,  Via Cintia, I-80126 Napoli, Italy}
\affiliation{INFN Sezione di Napoli, I-80126 Napoli, Italy}

\author{Thorsten Emig}
\affiliation{Laboratoire de Physique
Th\'eorique et Mod\`eles Statistiques, CNRS UMR 8626, B\^at.~100,
Universit\'e Paris-Sud, 91405 Orsay cedex, France}
\affiliation{Massachusetts Institute of Technology, MultiScale Materials Science
for Energy and Environment, Joint MIT-CNRS Laboratory (UMI 3466),
Cambridge, Massachusetts 02139, USA}
\affiliation{Massachusetts Institute of
Technology, Department of Physics, Cambridge, Massachusetts 02139, USA}

\author{Mehran Kardar}  
\affiliation{Massachusetts Institute of
Technology, Department of Physics, Cambridge, Massachusetts 02139, USA}

\begin{abstract}
The Casimir--Polder interaction between an anisotropic particle and a surface is orientation dependent. We study novel orientational effects that arise due to curvature of the surface for distances much smaller than the radii of curvature by employing a derivative expansion. 
For nanoparticles we derive a general short distance expansion of the interaction potential
in terms of their dipolar polarizabilities.  
Explicit results are presented for nano-spheroids made of SiO$_2$ and gold, both at zero and at finite temperatures. 
The preferred orientation of the particle is strongly dependent on curvature, temperature, as well as material properties.
\end{abstract}


\maketitle

\section{Introduction}

The interaction of small particles with surfaces is important to a plethora of phenomena in physics, chemistry and biology. While the cause of the interaction can differ, in many situations the particles are neutral (source free) and their interaction with the surface is due to an embedding fluctuating medium or field. There is considerable interest in investigating how the interaction is affected by the geometrical shape of the surface, and several experiments have~\cite{exp1,exp2,exp3,exp4}  probed dispersion forces between particles and micro-structured surfaces.  
More elaborate examples for this type of interaction include quantum frictional forces acting on particles  
moving along a surface~\cite{P1997}, the heat transfer between nanoparticles and curved or rough surfaces~\cite{BG2010}, and critical Casimir forces in colloidal systems and superfluid helium~\cite{KHD2009,VED2013}.

Here we consider forces induced by quantum (and thermal) fluctuations of the electromagnetic (EM) field, known as van der Waals or Casimir--Polder interactions. 
The forces between a particle (atom) and a flat surfaces have  been extensively studied~\cite{polder}; for recent reviews see~\cite{rev1,rev2}. 
However, roughness and curvature which are ubiquitous features of many surfaces
modify fluctuation--induced forces. 
Computing such interactions is complicated by their characteristic non-additivity 
which leads to interesting effects for anisotropic particles~\cite{MAP2015}. 
For good conductors, the force between spheroids scales not with the product of their actual volumes 
but with the product of the volumes of the enclosing spheres~\cite{EGJK2009}.  
The classic result of Balian and Duplantier~\cite{BD1977} and more recently developed scattering techniques~\cite{sca1,sca2} have been most successfully applied at distances that are large compared to the radii of curvature of the surface, and for a few specific surface shapes. 
A perturbative approach is presented in~\cite{messina}, where surfaces with smooth corrugations
of small amplitude, were studied. The validity of the latter is limited to particle--surface separations 
much larger than the corrugation amplitude. 

However, the regime most relevant to experiments is at short distances (compared to the radii of curvature of the surface). Analytical results are known only for specific geometries, like for a perfectly conducting cylinder and an atom~\cite{galina}.
A commonly used method in this regime is the proximity force approximation (PFA)~\cite{deri},
based on integrating the force to a flat plate over varying separations.
This approximation clearly fails for  anisotropic particles whose preferred orientation  depends
on the shape of the nearby surface.
Here, we employ a systematic approach that becomes exact in the limit of small particle--surface separations. It is based on an expansion of the interaction potential in derivatives of the surface profile, and hence applies to general, curved surfaces. An analogous  expansion has been used 
recently~\cite{fosco2,bimonte3,bimonte4} to study the Casimir interaction between two non-planar  surfaces. It has also been applied to other problems involving short range interactions between surfaces, like radiative heat transfer~\cite{golyk}, and stray electrostatic forces between conductors~\cite{fosco3}. 

The paper is organized as follows: In Sec. II we present the derivative expansion for the general case of a particle with electric and magnetic dipolar polarizabilities in front of a dielectric curved surface, and we specialize the results to the perfectly reflecting limit.  In Sec. III we compute explicitly the orientation dependence of the interaction for spheroids made of SiO$_2$ and gold, both at zero and at finite temperatures.  Section IV summarizes our results and provides an outlook.

\section{Derivative expansion of the Casimir-Polder potential}

\begin{figure} 
\includegraphics [width=.9\columnwidth]{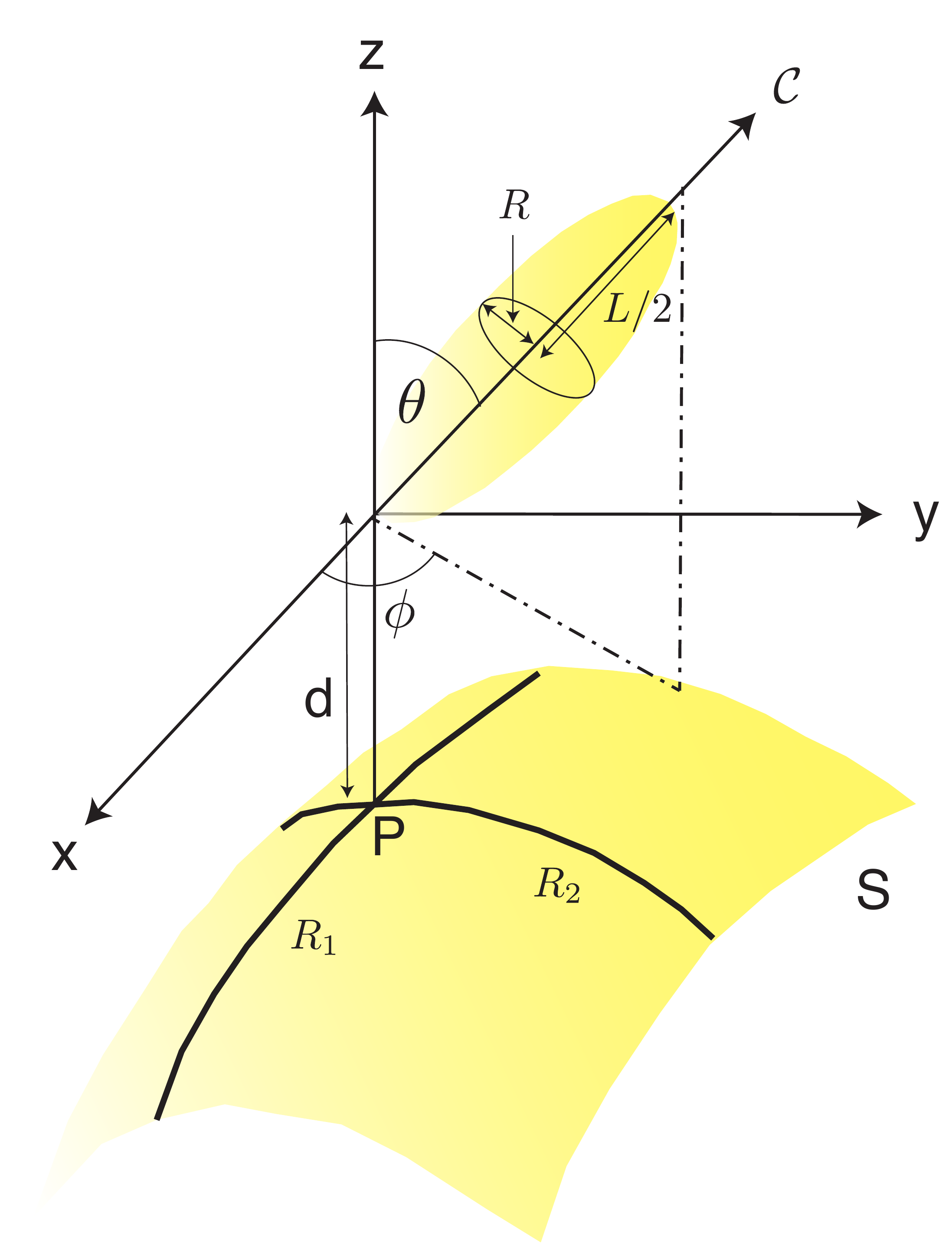} 
\caption{\label{fig1} Parametrization  the configuration   of  a nano-spheroid near a gently curved surface: Solid curves on the surface $S$ indicate the principal directions  at $P$ with local radii of curvature $R_1$ and $R_2$. Positive (negative) $R_j$ correspond to a surface that curves away from (towards) the particle.}  
\end{figure}

Consider a nanoparticle near a dielectric surface $S$. We assume that the particle is small enough (compared to the scale of its
separation $d$ to the surface) to be be considered as point-like, with its response to the electromagnetic 
fields fully described by the electric and magnetic dipolar polarizability tensors $\alpha^E_{\mu \nu}(\omega)$ and $\alpha^M_{\mu \nu}(\omega)$, respectively.  Let us denote by $\Sigma_1$ the plane through the particle which is orthogonal to the distance vector (which we take to be the ${\hat {\bf z}}$ axis) connecting the particle to the point $P$ of $S$ closest to the particle.  We assume that the surface $S$ is characterized by a {\it smooth} profile $z=H({\bf x})$, where ${\bf x} =(x,y)$ is the vector spanning $\Sigma_1$ (see Fig.~\ref{fig1}). In what follows Greek indices $\mu, \nu, \cdots$ label all coordinates $(x,y,z)$, while latin indices $i,j,k, \cdots$ refer to $(x,y)$ coordinates in the plane $\Sigma_1$. Throughout we adopt the convention that repeated indices are summed over.  

The exact Casimir-Polder potential at finite temperature $T$  is given by the   formula~\cite{sca1,sca2}
\begin{equation}
\label{eq.4}
U= - k_B T \sideset{}{'}\sum_{n=0}^\infty \text{Tr} \, [ {\mathbb T}^{(S)} {\mathbb U} {\mathbb T}^{(P)}  {\mathbb U}  ](\kappa_n)\;.
\end{equation} 
Here ${\mathbb T}^{(S)}$ and ${\mathbb T}^{(P)}$  denote, respectively, the scattering T-operators of the plate $S$ and the particle, evaluated  at the Matsubara wave numbers $\kappa_n=2\pi n k_BT/(\hbar c)$, and the primed sum indicates that the $n=0$ term carries  weight $1/2$.   In a plane-wave basis $|{\bf k},Q \rangle$~\cite{fn3} where ${\bf k}$ is the in-plane wave-vector, and $Q=E,M$ labels respectively electric (transverse magnetic) and magnetic (transverse electric) modes,
the translation operator $\mathbb U$ in Eq.~(\ref{eq.4}) is diagonal with matrix elements
$e^{-d q}$ where
$q=\sqrt{{ k}^2+ \kappa_n^2}\equiv q(k)$, $k=|{\bf k}|$. The matrix elements of the particle  T-operator in dipole approximation are 
\begin{align}
 { {\cal T}}_{QQ'}^{(P)}({\bf k}, {\bf k}' ) & =-\frac{2 \pi \kappa_n^2}{ \sqrt{q q'}}  \left( e^{(+)}_{Q \mu}({\bf k}) \alpha^E_{\mu \nu}({\rm i} c\kappa_n) e^{(-)}_{Q' \nu}({\bf k}') \right. \nonumber \\ 
&
+ \left. {\tilde e}^{(+)}_{Q \mu}({\bf k}) \alpha^M_{\mu \nu}({\rm i} c\kappa_n) {\tilde e}^{(-)}_{Q' \nu}({\bf k}') \right)\,,
\end{align}
where $q'=q(k')$, ${\bf e}^{(\pm)}_{M}({\bf k})={\hat {\bf z}} \times {\hat {\bf k} }$, ${\bf e}^{(\pm)}_{E}({\bf k})=-1/\kappa_n ({\rm i} k {\hat {\bf z}} \pm q  {\hat {\bf k} })$,  ${\hat {\bf k} }={\bf k}/k$ and  we set ${\tilde {\bf e}}_E^{(\pm)}=-{\bf e}_M^{(\pm)}$,  ${\tilde {\bf e}}_M^{(\pm)}={\bf e}_E^{(\pm)}$.
The T-operator  ${\mathbb T}^{(S)}$ of an arbitrary curved plate is not known in closed form, and its computation is in general quite challenging, even numerically.  
In Ref.~\cite{BEK2014}, however, the leading curvature corrections to the potential
were computed for an atom in front of a smoothly curved surface,
in the experimentally relevant limit of small separations. 
The key idea is that as the Casimir--Polder interaction falls off rapidly with separation, 
it is  reasonably expected that the potential $U$ is dominated by 
a small neighborhood of the point $P$ of $S$ which is {\it closest} to the particle. This physically plausible idea suggests that for small separations $d$, the potential $U$ can be expanded as a series in an increasing number of derivatives of the height profile $H$, evaluated at the particle's position. 
Up to fourth order, and assuming that the surface is homogeneous and isotropic, the most general expression which is invariant under rotations of the $(x,y)$ coordinates, and that involves at most four derivatives of $H$ (but no first derivatives since $\nabla H ({\bf 0})=0$) can be expressed (up to ${\cal O}(d^{-1}$)) as 
\begin{widetext}
\begin{align}
\label{derexpa}
  U & = -\frac{k_B T}{d^3} \sideset{}{'}\sum_{n=0}^\infty \sum_{P=E,M}\left\{\beta^{(0)}_{P|1} \alpha^{P}_\perp + \beta^{(0)}_{P|2} \alpha_{zz}^{P} + d\times\left[ (\beta^{(2)}_{P|1} \alpha_\perp^{P} + \beta^{(2)}_{P|2} \alpha_{zz}^{P}) \nabla^2 H+
 \beta^{(2)}_{P|3}  \left(\partial_i \partial_j H - \frac{1}{2} \nabla^2 H \delta_{ij}\right) \alpha_{ij}^{P}\right] \right. \nonumber \\ 
& 
+d^2 \times \left[\frac{}{}\beta^{(3)}_P  \alpha_{zi}^{P} \partial_i \nabla^2 H 
+(\nabla^2 H)^2 (\beta^{(4)}_{P|1} \alpha_\perp^{P} + \beta^{(4)}_{P|2} \alpha_{zz}^{P} ) +  (\partial_i \partial_j H)^2 (\beta^{(4)}_{P|3} \alpha_\perp^{P}\! + \beta^{(4)}_{P|4} \alpha_{zz}^{P} )\right.\nonumber \\
&
\left. \left.+ \beta^{(4)}_{P|5}
\nabla^2 H \! \left(\partial_i \partial_j H - \frac{1}{2} \nabla^2 H \delta_{ij}\right) \alpha_{ij}^{P} \right]\right\}\,,
\end{align}
\end{widetext}
where $\alpha^{P}_\perp=\alpha_{xx}^{P}+\alpha_{yy}^{P}$, and it is understood that all derivatives of $H({\bf x})$ are evaluated at the particle's position, i.e., for ${\bf x}={\bf 0}$. The coefficients
$\beta^{(p)}_{P|q}$ are dimensionless functions of $\xi_n=2\pi n k_B T/(\hbar c)$, and of any other dimensionless ratio of frequencies characterizing the material of the surface. The derivative expansion  in Eq.~(\ref{derexpa}) can be formally obtained by a re-summation of the perturbative series for the potential for small in-plane momenta ${\bf k}$~\cite{BEK2014}. We note that there are additional terms involving four derivatives of $H$ which, however, yield contributions $\sim 1/d$ (as do terms involving five derivatives of $H$) and are hence  neglected.

A geometrical interpretation of Eq.~(\ref{derexpa}) is obtained when the $x$ and $y$ axis are chosen to  coincide with the principal directions of curvature of $S$ at $P$.  Then the expansion of $H$ 
is $H=d+x^2/(2 R_1)+ y^2/(2 R_2)+\cdots$, where $R_1$ and $R_2$ are the radii of curvature at $P$.  In this coordinate system, the derivative expansion of $U$ reads 
\begin{widetext} 
\begin{align} 
\label{derexpa2}
  U & =-\frac{k_B T}{d^3} \sideset{}{'}\sum_{n=0}^\infty\sum_{P=E,M}\left\{ \beta^{(0)}_{P|1} \alpha^{P}_\perp  + \beta^{(0)}_{P|2} \alpha_{zz}^{P}+ \left(\frac{d}{R_1}+\frac{d}{R_2} \right)  (\beta^{(2)}_{P|1} \alpha_\perp^{P} + \beta^{(2)}_{P|2} \alpha_{zz}^{P})+
 \frac{\beta^{(2)}_{P|3}}{2}  \left(\frac{d}{R_1}-\frac{d}{R_2}\right) (\alpha^P_{xx}-\alpha^P_{yy}) \right.\nonumber \\
&+ d^2 \beta^{(3)}_P  \alpha^P_{zi} \partial_i \left(\frac{1}{R_1}
+\frac{1}{R_2} \right) 
+ \left(\frac{d}{R_1}+\frac{d}{R_2} \right)^2  (\beta^{(4)}_{P|1} \alpha_\perp^{P} + \beta^{(4)}_{P|2} \alpha_{zz}^{P} )\frac{}{}
\nonumber \\
& \left.\left.
+ \left[\left(\frac{d}{R_1}\right)^2+\left(\frac{d}{R_2} \right)^2 \right]    (\beta^{(4)}_{P|3} \alpha_\perp^{P}\! + \beta^{(4)}_{P|4} \alpha_{zz}^{P} )+ \frac{\beta^{(4)}_{P|5}}{2} \left[\left(\frac{d}{R_1}\right)^2-\left(\frac{d}{R_2} \right)^2  \right] (\alpha^P_{xx}-\alpha^P_{yy}) 
  \right\}\right.\;.
\end{align}
\end{widetext}

As demonstrated in Ref.~\cite{BEK2014}, the coefficients $\beta^{(p)}_{P|q}$ in Eq.~(\ref{derexpa}) can be extracted from  
the perturbative series of the potential $U$. To 
second order in the deformation $h({\bf x})=H({\bf x})-d$, this involves an expansion of the T-operator of the surface $S$ to the same order.
The latter expansion was obtained in Ref.~\cite{voron} for a dielectric material described by a frequency dependent permittivity $\epsilon(\omega)$. It reads
\begin{align}
& { {\cal T}}_{QQ'}^{(S)}({\bf k}, {\bf k}' )=(2 \pi)^2 \delta^{(2)}({\bf k}-{\bf k'})\,\delta_{QQ'}\, r^{(S)}_{Q} (i c\kappa_n,{\bf k})
\nonumber\\
&+ \sqrt{q\,q'}\,\left[-2 \,B_{QQ'}({\bf k}, {\bf k}')\,\tilde{h}({{\bf k}- {\bf k}'})\right. \\
 & \left. + \!\!\int \!\!\frac{d^2 {\bf k}''}{(2 \pi)^2} (B_2)_{QQ'}({\bf k}, {\bf k}';{\bf k}'') \tilde{h}({{\bf k}\!- \!{\bf k}''}) \tilde{h}({{\bf k}''\!-\! {\bf k}'})+\dots \right]\;, \nonumber
\end{align}
where $r^{(S)}_{Q} (i c\kappa_n,{\bf k})$ denote the familiar Fresnel reflection coeffcients of a flat surface, and $\tilde{h}({\bf k})$ is the Fourier transfromed deformation.  Explicit expressions for the kernels $B_{Q Q'}({\bf k}, {\bf k}')$ and $(B_2)_{Q Q}({\bf k}', {\bf k}';{\bf k}'')$ are given in Ref.~\cite{voron}. Computing the  coefficients $\beta^{(p)}_{P|q}$ involves an integral over ${\bf k}$ and ${\bf k'}$ (as it is apparent from Eq.~(\ref{eq.4})) that cannot be performed analytically for a dielectric plate. In the following, we shall consider a perfect conductor, in which case the integrals can be carried out analytically.  In this case, the matrix $B_{QQ'}({\bf k}, {\bf k}')$ takes the simple form
\be
B({\bf k}, {\bf k}')=\left(\begin{array}{cc} \frac{{\hat{\bf k}}\cdot{\hat {\bf k}}'  \kappa_n^2+k k'}{ q q'}& \frac{\kappa_n}{ q} {\hat {\bf z}}\cdot({\hat{\bf k}}\times {\hat {\bf k}}') \\ \frac{\kappa_n}{ q'} {\hat {\bf z}}\cdot({\hat{\bf k}}\times {\hat {\bf k}}') & -{\hat{\bf k}}\cdot{\hat {\bf k}}' \\ \end{array} \right)\;,
\ee 
where the matrix entries $Q,\,Q'$ correspond to $E,M$ respectively.  Also, the matrix $(B_2)_{QQ'}({\bf k}, {\bf k}';{\bf k}'') $ is simply related to $B$ by
\be
(B_2)({\bf k}, {\bf k}';{\bf k}'') =2 q'' B({\bf k}, {\bf k}'') \sigma_3 B({\bf k}'', {\bf k}')\;,
\ee
where $\sigma_3={\rm diag}(1,-1)$. The coefficients $\beta^{(p)}_{P|q}$ are now functions of $\xi$ only, and we list them in  Tables~\ref{tab:betas}, \ref{tab:betasmag} for electric and magnetic dipole polarizabilities, respectively. 

\begin{table*}
\begin{tabular}{|c|c|l|l|}
\hline
p & q & $\times e^{-2\xi}$  & $\times \text{Ei}(2\xi)$ \\
\hline
0 & 1 &  $\frac{1}{8}(1+2 { \xi}+4 { \xi}^2)$ &  $0$\\
 & 2 & $\frac{1}{4}(1+2 { \xi} )$ & $0$\\
2 & 1 & $-\frac{1}{32}(3+6 { \xi} +6 { \xi}^2+4 { \xi}^3 )$ & $-\frac{ { \xi}^4}{4} $\\
 & 2 & $-\frac{1}{16}(1+2 { \xi} -2 { \xi}^2+4 { \xi}^3 )$ & ${ \xi}^2 \left(1-\frac{{ \xi}^2}{2} \right)$\\
 & 3 & $-\frac{1}{32}(3+6 { \xi} +2{ \xi}^2-4 { \xi}^3 )$  & $\frac{ { \xi}^4}{4}$\\
3 & & $\frac{1}{32}(1+2 { \xi} -2{ \xi}^2+4 { \xi}^3 )$ & $-\frac{ { \xi}^2}{4} (2-{ \xi}^2)$\\
4 & 1 & $\frac{1}{384}(3+6 { \xi} +15{ \xi}^2+22 { \xi}^3+2 { \xi}^4-4{ \xi}^5 )$ & $\frac{ { \xi}^4}{48} (6-{ \xi}^2)$\\
 & 2 &  $-\frac{1}{960}(15+542 { \xi} +259{ \xi}^2-546 { \xi}^3-14 { \xi}^4+28{ \xi}^5 )$ & $- 2 \xi^2 (1-\frac{7 \xi^2}{12}+\frac{7 \xi^4}{240})$\\
 & 3 &  $\frac{1}{192}(15+30 { \xi} -9{ \xi}^2+70 { \xi}^3+2 { \xi}^4-4{ \xi}^5 )$ & $\frac{ { \xi}^4}{24} (18-{ \xi}^2) $\\
 & 4 &  $\frac{1}{480}(45+218 { \xi} -59{ \xi}^2+146 { \xi}^3+14 { \xi}^4-28{ \xi}^5 )$ & $\frac{ { \xi}^4}{60} (40-7{ \xi}^2)$\\
 & 5 &  $\frac{1}{96}(9+18 { \xi} -27{ \xi}^2+50 { \xi}^3-2 { \xi}^4+4{ \xi}^5 )$ & $ { \xi}^4\left(1+\frac{{ \xi}^2}{12}\right)$\\
\hline
\end{tabular}
\caption{\label{tab:betas} The coefficients $\beta^{(p)}_{E|q}$ for the electric dipole contribution are obtained by multiplying the third column by $e^{-2\xi}$, and adding the fourth column times $\text{Ei}(2\xi)=-\int_{2\xi}^\infty dt \exp(-t)/t$.}
\end{table*}

\begin{table*}
\begin{tabular}{|c|c|l|l|}
\hline
p & q & $\times e^{-2\xi}$  & $\times \text{Ei}(2\xi)$ \\
\hline
0 & 1 &  $-\frac{1}{8}(1+2 { \xi}+4 { \xi}^2)$ &  $0$\\
 & 2 & $-\frac{1}{4}(1+2 { \xi} )$ & $0$\\
2 & 1 & $\frac{1}{32}(5+10 { \xi} +10 { \xi}^2-4 { \xi}^3 )$ & $\frac{ { \xi}^2}{2} \left(1-\frac{\xi^2}{2} \right) $\\
 & 2 & $\frac{1}{16}(3+6 { \xi} +2 { \xi}^2-4 { \xi}^3 )$ & $-\frac{ \xi^4}{2}  $\\
 & 3 & $\frac{1}{32}(1+2 { \xi} -2{ \xi}^2+4 { \xi}^3 )$  & $\frac{3 { \xi}^2}{2} \left(1+\frac{\xi^2}{6} \right)$\\
3 & & $\frac{1}{32}(5+10 { \xi} -2{ \xi}^2+4 { \xi}^3 )$ & $\frac{ { \xi}^2}{4} (4+{ \xi}^2)$\\
4 & 1 & $-\frac{1}{960}(165-438 { \xi} +339{ \xi}^2-466 { \xi}^3-14 { \xi}^4+28{ \xi}^5 )$ & $\frac{ { \xi}^2}{2} (1+2{ \xi}^2-\frac{7 \xi^4}{60})$\\
 & 2 &  $-\frac{1}{192}(15+30 { \xi} +9{ \xi}^2-22 { \xi}^3-2 { \xi}^4+4{ \xi}^5 )$ & $\frac{ {\xi}^4}{4} \left(1-\frac{ \xi^2}{6} \right)$\\
 & 3 &  $-\frac{1}{960}(105+722 { \xi} +139{ \xi}^2-66 { \xi}^3-14 { \xi}^4+28{ \xi}^5 )$ & $-\frac{ {3 \xi}^2}{2} \left(1-\frac{\xi^2}{9}+\frac{7 \xi^4}{180}\right) $\\
 & 4 &  $-\frac{1}{96}(3+6 { \xi} +33{ \xi}^2-70 { \xi}^3-2 { \xi}^4+4{ \xi}^5 )$ & $\frac{ {3 \xi}^4}{2} \left(1-\frac{ \xi^2}{18}\right)$\\
 & 5 &  $-\frac{1}{480}(15+158 { \xi} +121{ \xi}^2-214 { \xi}^3+14 { \xi}^4-28{ \xi}^5 )$ & $ -\frac{5 \xi^2}{2}\left(1-\frac{{ \xi}^2}{3}-\frac{7 \xi^4}{150}\right)$\\
\hline
\end{tabular}
\caption{\label{tab:betasmag} The coefficients $\beta^{(p)}_{M|q}$ for the magnetic dipole contribution, using the same notation as in Tab.~\ref{tab:betas}.}
\end{table*}

\section{Orientation dependence}

In this section we investigate the shape and orientation dependence of the Casimir--Polder force using Eq.~(\ref{derexpa2}).  Before we consider a curved surface, it is 
interesting to stress that for a perfectly reflecting planar surface there is {\it no} orientation dependence at zero temperature for dipolar particles with frequency independent polarizabilities as realized, e.g., in the perfectly conducting limit. This follows directly from the fact that the $\xi$ integrals of the two coefficients $\beta^{(0)}_{P|1}$ and $\beta^{(0)}_{P|2}$ are equal so that the potential is proportional to the rotationally invariant trace of $\alpha^E - \alpha^M$~\cite{EGJK2009}. Coming back to a curved surface, we
assume for simplicity that its height profile $H$ is invariant under  independent reflections in the $x$ and $y$ directions. This symmetry of the surface ensures that the term proportional to $\beta^{(3)}_P$ in Eq.~(\ref{derexpa2}) is absent. Moreover, we assume that the particle has one axis of rotational symmetry. In a new orthogonal basis $(1,2,3)$ oriented such that the third axis coincides with the particle's symmetry axis ${\cal C}$, the polarizability tensors   are diagonal with  $\tilde{\alpha}^P ={\rm diag} (\tilde{\alpha}^P_{\perp}/2,\tilde{\alpha}^P_{\perp}/2,\tilde{\alpha}^P_{33} )$.

\begin{figure}[th]
\includegraphics [width=.9\columnwidth]{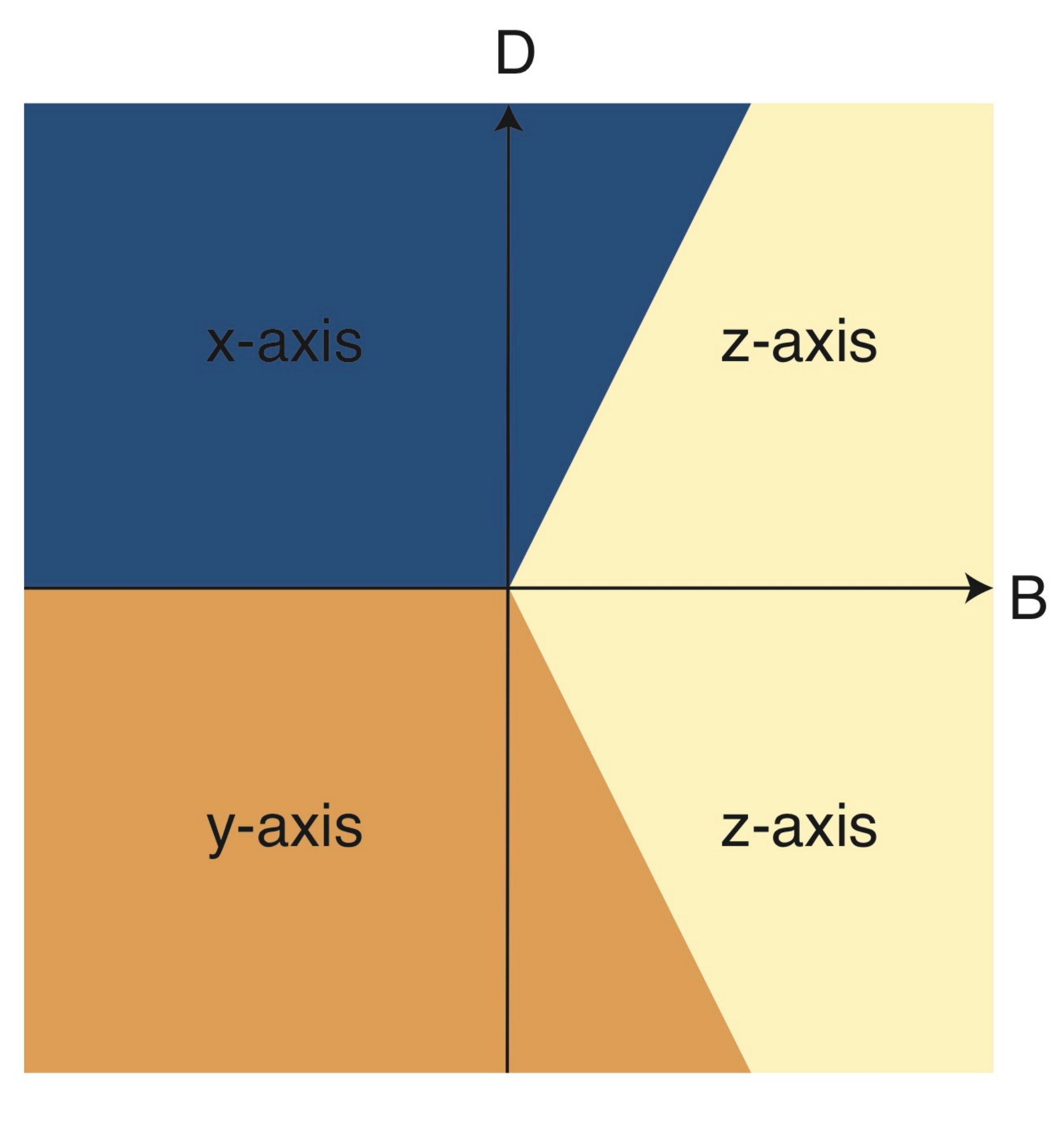} 
\caption{\label{fig:stability} Stable orientations of the particle's symmetry axis as function of the coefficients $B$ and $D=C(d/R_1-d/R_2)$ in Eq.~\eqref{potang}.}
\end{figure}

\begin{figure}[th]
\includegraphics [width=1.0\columnwidth]{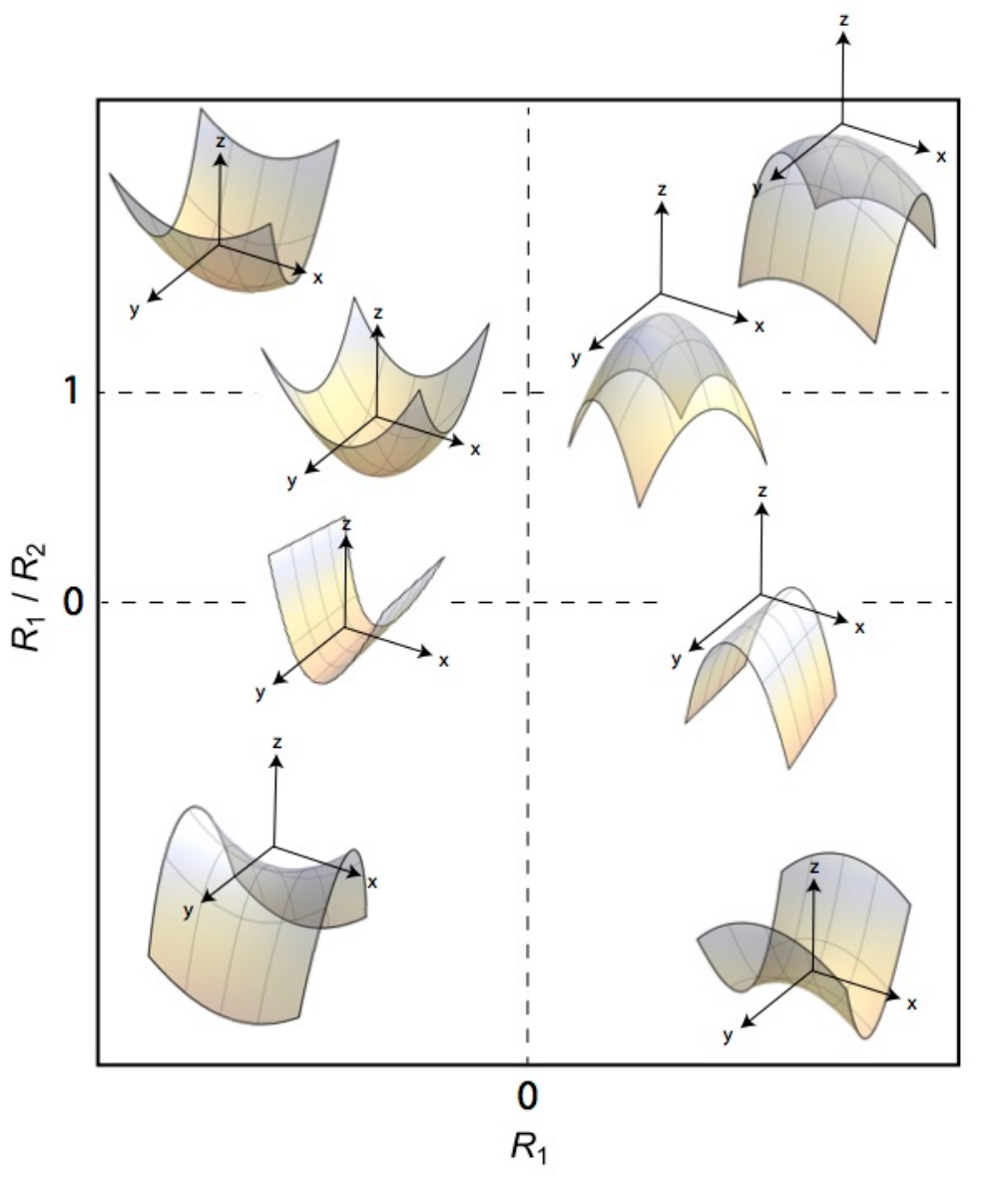} 
\caption{\label{fig:shapes} Typical surface shapes for limiting cases corresponding to combinations of $R_1=\pm 1$, $R_2=\pm 3$ and $R_2\to\infty$ (arbitrary units). The coordinate frames indicate the position and orientation of the particle.}
\end{figure}

The polarizability tensors for an arbitrary orientation are then obtained as $\alpha=  {\cal R}^{-1} \tilde{\alpha} {\cal R}$, where ${\cal R}$ is the  matrix that rotates the principal axis of the particle to the basis composed of the principal directions of the surface $S$, i.e. ${\cal R}(1,2,3) \rightarrow (x,y,z)$. The orientation of the particle is conveniently parametrized by the polar angles $(\theta, \phi)$ of its symmetry axis ${\cal C}$, where $\theta$ is the angle  formed by   ${\cal C}$ and the $z$ axis,  and $\phi$ is the angle between the $(z,{\cal C})$ plane and the $(x,z)$ plane (see Fig.~\ref{fig1}). The polarizability tensors in the two coordinate systems are related by
\be
\alpha^P_{\perp}=\frac{1}{4}[3\, \tilde{\alpha}^P_\perp +2\, \tilde{\alpha}^P_{33}-{\sigma} ^P \cos(2 \theta)]\;,\label{alpha1}
\ee
\be
\alpha^P_{zz}=\frac{1}{4}[  \tilde{\alpha}_\perp^P+2 \,\tilde{\alpha}_{33}^P + {\sigma}^P\cos(2 \theta)]\;,
\ee
\be
\alpha^P_{xx}-\alpha^P_{yy}=\frac{{\sigma}^P }{2}  \cos(2 \phi) \sin^2 \theta\;,\label{alpha3}
\ee
where we defined ${\sigma}^P=2  \,\tilde{\alpha}_{33}^P- \tilde{\alpha}_\perp^P$. Since the $x$ and $y$ axis are chosen to  coincide with the principal directions of $S$ at $P$, Eqs.~(\ref{alpha1}-\ref{alpha3}) together with 
Eq.~(\ref{derexpa2}) yield the potential in the simple form 
\begin{widetext}
\begin{equation}
U=-\frac{k_B T \, V}{ d^3}\left[A(d)+ B(d) \cos(2 \theta)+ C(d)\;\left(\frac{d}{R_1}-\frac{d}{R_2}\right)\cos(2 \phi) \sin^2(\theta) \right]\,,
\label{potang}
\end{equation}
\end{widetext}
where $A(d),\, B(d),\, C(d)$ are in general functions of temperature and the ratios $d/R_i$, but do not depend on the angles $\theta$ and $\phi$, and $V$ is the volume of the particle. Before turning to detailed computations, we briefly discuss the qualitative features of the potential. It is obvious that the coefficients $B$ and $C$ must vanish for a spherical particle. For a non-spherical particle in front of a planar surface ($R_1=R_2\to\infty$) the potential  $U$ in Eq.~(\ref{potang}) is invariant under rotations about the $z$-axis, as expected. The coefficient  $B$, however, is in general different from zero, and hence even for a planar surface  the potential depends on the polar angle $\theta$ (except, as discussed above, for a perfectly reflecting surface and for frequency independent polarizabilities).  In order to have  a non-trivial dependence of $U$ on the azimuthal angle $\phi$, it is necessary to break the rotational symmetry about the $z$-axis. Clearly, this  happens when the surface has different radii of curvature $R_1 \neq R_2$ at $P$, as evidenced by the third term between the brackets of Eq.~(\ref{potang}). For $R_1 \neq R_2$,
it is  easy to verify that in general the potential $U$  has a {\it unique} minimum,  corresponding to an orientation of the particle along one of the $(x,y,z)$ axes. More precisely, with $D\equiv C(d/R_1-d/R_2)$, the stable orientation of the particle's symmetry axis is along the
\begin{enumerate}
\item[(i)]  $x$ axis if  $D>\max\{0,2B\}$,
\item[(ii)]  $y$ axis if  $D<\min\{0,-2B\}$, 
\item[(iii)]  $z$ axis otherwise.
\end{enumerate}
The stable orientations are summarized in the diagram of Fig.~\ref{fig:stability}. To grasp more easily the different orientations of the particle relative to the curved surface, we show in Fig.~\ref{fig:shapes}
the typical surface shapes for positive and negative radii of curvature, along with the coordinate frames for the position and orientation of the particle. 


Below, we  numerically compute the potential between  a gold surface and a spheroidal particle, made either of gold or of  vitreous ${\rm Si O}_2$. For particle--surface separations $d$ larger than the plasma wavelength of gold, $\lambda_P=2 \pi c/\omega_p \simeq 120$nm, and smaller than a few micron, as we shall consider, the penetration depth of the electromagnetic fields in gold contributing to the Casimir-Polder potential is  $\delta_{\rm gold} \le$ 20nm, and therefore it is always much smaller than the separation $d$. In this range of separations, the gold surface $S$  can  thus be considered as  perfectly reflecting, and it is therefore  justified to use in Eq.~(\ref{derexpa2})  the expressions  of the $\beta_{P|q}^{(p)}$ coefficients  for a perfect  conductor, which are listed in Tables I and II.

\subsection{${\rm SiO}_2$ particle}   

For a dielectric ellipsoid with electric permittivity $\epsilon$  (and magnetic permeability $\mu=1$), the polarizability tensor $\alpha^{E}$ is diagonal with respect to its principal axes, with elements (for $\mu \in \{1, 2, 3\}$)
\be
\tilde{\alpha}_{\mu \mu}^E = \frac{V}{4 \pi} \frac{\epsilon -1}{1+(\epsilon-1) n_{\mu}}\;,\label{alphaE}
\ee
where $V = 4 \pi r_1 r_2 r_3/3$ is the ellipsoid's volume. In the case of spheroids, for which $r_1 = r_2 = R$ and $r_3 = L/2$, the so-called depolarizing factors can be expressed in terms of elementary functions,
\begin{align}
&n_1=n_2=\frac{1-n_3}{2}\;,\nonumber\\
&n_3=\frac{1-e^2}{2 e^3}\left(\log \frac{1+e}{1-e} - 2 e\right)\;,
\end{align}
where the eccentricity $e=\sqrt{1-4 R^2/L^2}$ is real for a prolate spheroid ($L > 2R$) and imaginary for an oblate spheroid ($L < 2R$). For a prolate spheroid $0<n_3 < 1/3 $, while for an oblate spheroid $1/3<n_3 < 1$, the value $n_3=1/3$ corresponding to a sphere.  For the dynamic permittivity $\epsilon({\rm i} \,\omega)$ along the imaginary frequency axis, we use the simple two-oscillator model \be \epsilon({\rm i} \,\omega)=1+\frac{C_{\rm UV} \omega^2_{\rm UV}}{\omega^2+\omega^2_{\rm UV}}+\frac{C_{\rm IR} \omega^2_{\rm IR}}{\omega^2+\omega^2_{\rm IR}}\;, \ee with the parameters $C_{\rm UV}=1.098$, $C_{\rm IR}=1.703$, $\omega_{\rm UV}=2.033 \times 10^{16}$ rad/s, and $\omega_{\rm IR}=1.88 \times 10^{14}$ rad/s, which were obtained by a fit to optical data for ${\rm SiO}_2$~\cite{hough} .

\begin{figure} 
\includegraphics [width=.9\columnwidth]{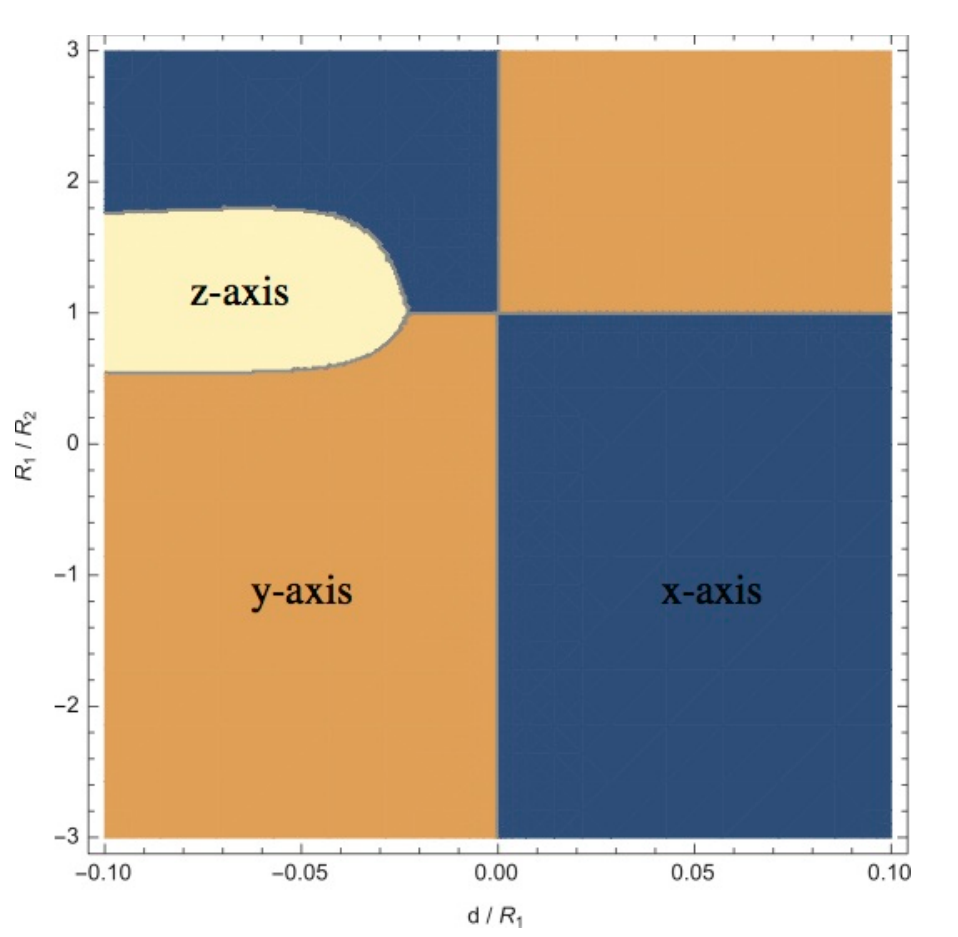} 
\caption{\label{fig_stability_SiO2_pancake} Stability diagram for a ${\rm SiO}_2$ oblate spheroid (``pancake'') with $n_3=0.7$ and $R_1=1000\mu$m at $T=0$ K.}
\end{figure}

\begin{figure} 
\includegraphics [width=.9\columnwidth]{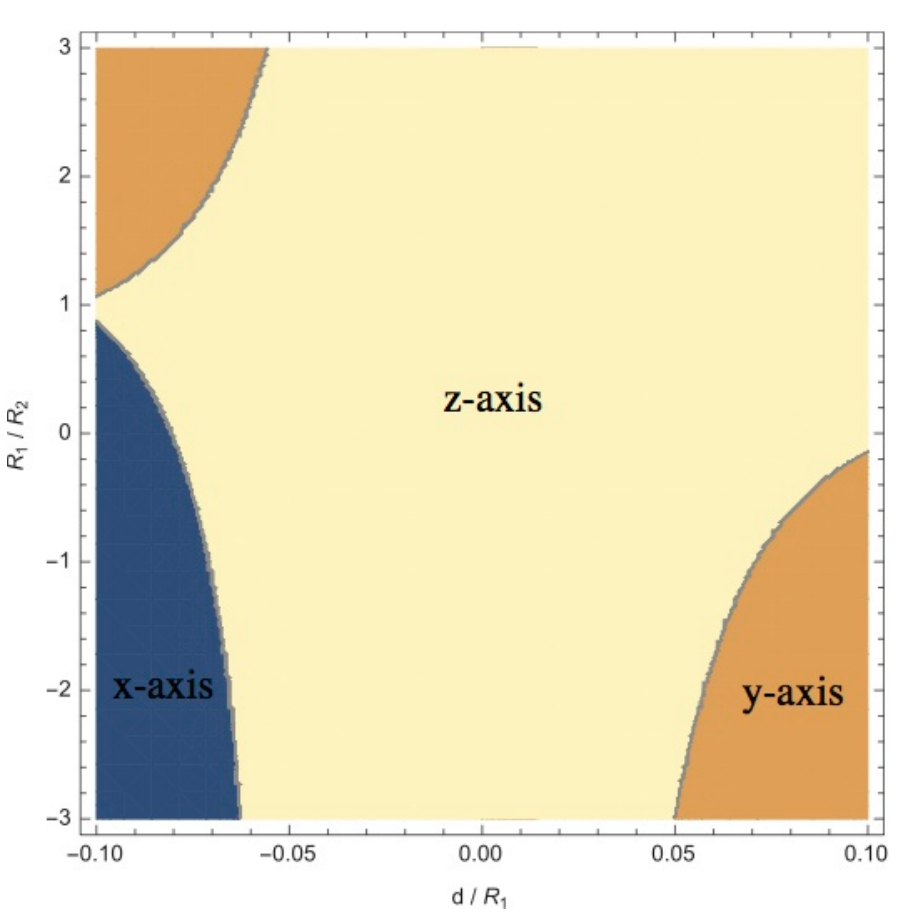} 
\caption{\label{fig_stability_SiO2_needle} Stability diagram for a ${\rm SiO}_2$ prolate spheroid (``needle'') with $n_3=0.2$ and $R_1=100 \, \mu$m at $T=0$ K.}
\end{figure}

We observed earlier that  the potential $U$ is minimized when the particle's axis points in the direction of one of the coordinate axes. First we determine the preferred orientations at zero temperature.
In Fig. \ref{fig_stability_SiO2_pancake} we show the stability diagram for a ${\rm SiO}_2$ oblate spheroid with $n_3=0.7$  (``pancake'') and $R_1=1000 \, \mu$m. An analogous diagram for a prolate spheroid with $n_3=0.2$ (``needle'')  and $R_1=100 \, \mu$m is show in Fig.~\ref{fig_stability_SiO2_needle}. From the diagrams it can be observed that the signs of surface curvature have an important effect on the preferred orientation of the particle. We note that the diagram depends on the choice of $R_1$ since it is compared to the material dependent length scales that are set by the characteristic frequencies of SiO$_2$. 

\begin{figure} 
\includegraphics [width=.9\columnwidth]{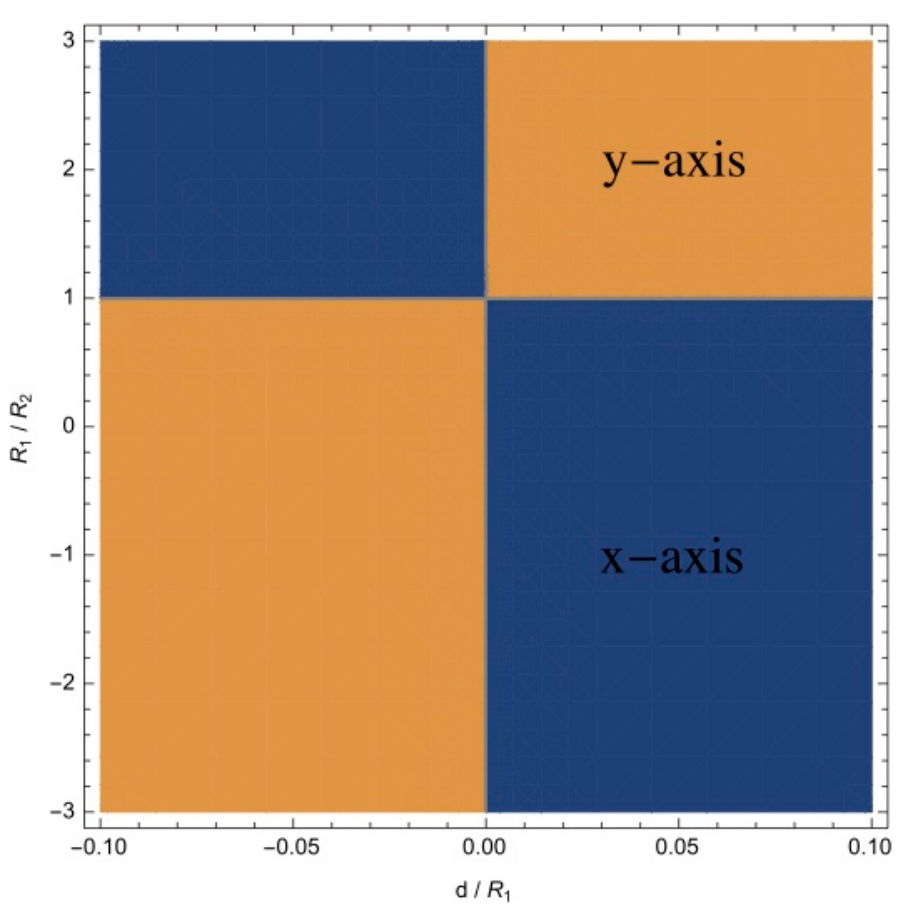} 
\caption{\label{fig_stability_SiO2_pancake_finiteT} Stability diagram for a ${\rm SiO}_2$ oblate spheroid (``pancake'') with $n_3=0.7$ and $R_1=1000 \, \mu$m at $T=300$ K.}
\end{figure}

Thermal fluctuations have a strong impact on the stable orientation of the particle. For room temperature, $T=300$K, the stability diagram for a ``pancake'' is shown in Fig.~\ref{fig_stability_SiO2_pancake_finiteT}. Only the $x$ and $y$ axes occur as stable directions, and the boundaries of the stable regions are simply given by $R_1=R_2$ and $R_1\to\infty$.  A ``needle'' at $T=300$ K is always oriented along the $z$-axis in the parameter range of the stability plots shown here.

\subsection{Gold particle}

\begin{figure}[t]
\includegraphics [width=.9\columnwidth]{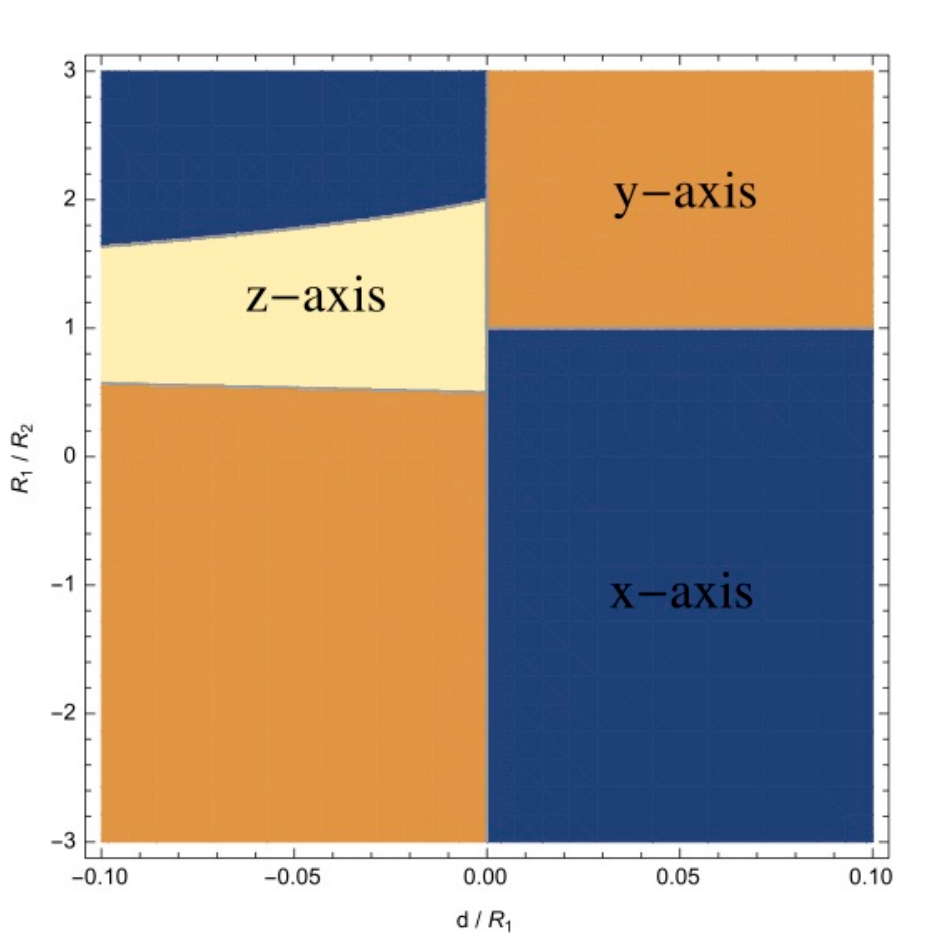} 
\caption{\label{fig_stability_Au_pancake} Stability diagram for a gold  oblate spheroid (``pancake'') with $n_3=0.7$ at $T=0$ K.}
\end{figure}

\begin{figure}[t]
\includegraphics [width=.9\columnwidth]{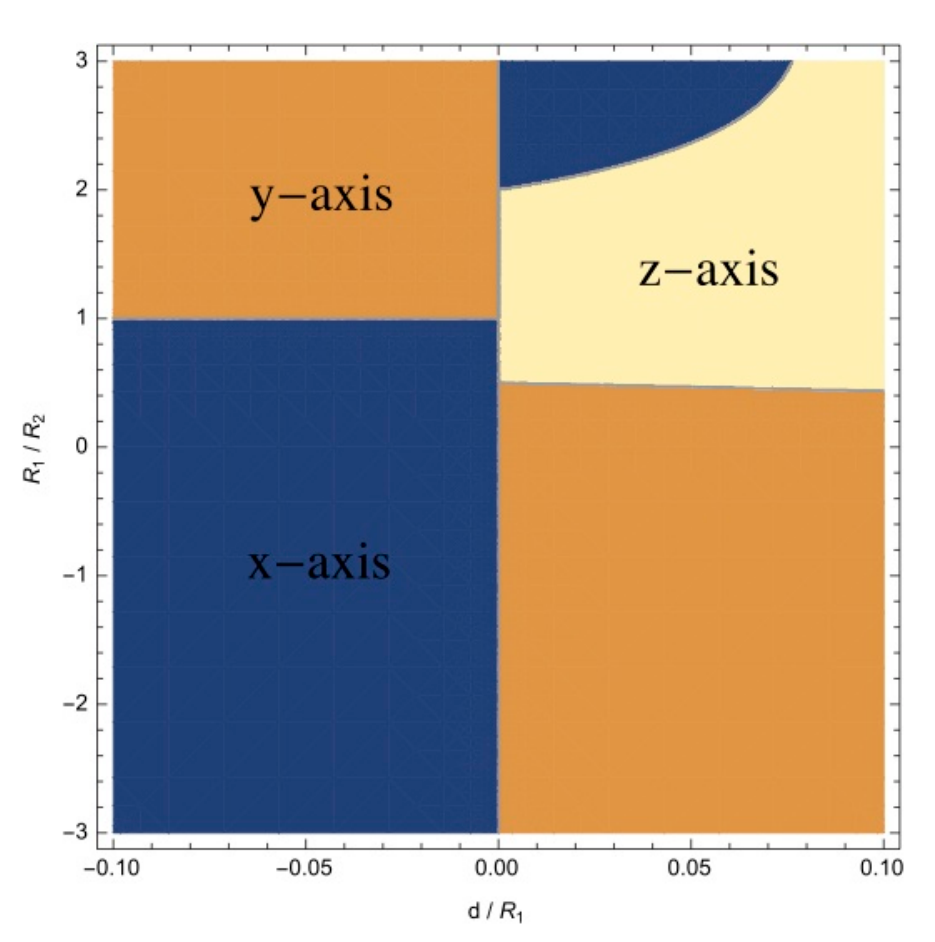} 
\caption{\label{fig_stability_Au_needle} Stability diagram for a gold prolate spheroid (``needle'') with $n_3=0.2$ at $T=0$ K.}
\end{figure}

Next we consider a gold spheroid at zero temperature. As explained above,  the penetration depth $\delta_{\rm gold}$ in gold of the electromagnetic fields that contribute to the potential $U$ is always less than $20$nm, for separations $d$ larger than $\lambda_P$ and less than a few microns. A nano-particle of characteristic size $\ell$, satisfying the condition $\delta_{\rm gold} \ll \ell \ll d$, can be modeled as perfectly reflecting. For such a particle, both the electric and magnetic dipolar polarizablities need to be considered.  The electric dipolar polarizability ${\tilde \alpha}^E$ is given by Eq.~(\ref{alphaE}) with $\epsilon\to\infty$.  The dipolar magnetic  polarizability  ${\tilde \alpha}^M$ coincides with  that of perfectly diamagnetic spheroid,  and can thus can be obtained by setting $\mu=0$ in the formula for the magnetic polarizability of a magnetizable spheroid, given by
\begin{equation}
\tilde{\alpha}_{\nu \nu}^M = \frac{V}{4 \pi} \frac{\mu -1}{1+(\mu-1) n_{\nu}}\;.
\label{alphaM}
\end{equation}
Since the dipolar polarizabilities $\alpha^P$ of a perfectly conducting particle are frequency independent, the frequency integrals in Eq.~(\ref{derexpa2}) can be performed analytically. At $T=0$, the potential is then given by the explicit expression
\begin{widetext} 
\begin{align} 
  U &  =-\frac{\hbar \, c \, V}{32 \pi^2 n_3\,(1-n_3^2)\, d^4}\left\{1+9 n_3 - \frac{17+183 n_3-14 n_3^2}{30}\left(\frac{d}{R_1}+\frac{d}{R_2} \right)+\frac{215+2457 n_3-434 n_3^2}{420} \left[\left(\frac{d}{R_1} \right)^2+\left(\frac{d}{R_2} \right)^2 \right]
 \right.  \nonumber \\
&+ \frac{11+693 n_3-266 n_3^2}{210} \frac{d^2}{R_1 R_2}+
 \frac{1-3 n_3}{30} \left\{ \left\{(1+2 n_3)\left(\frac{d}{R_1}+\frac{d}{R_2}\right)+\frac{23+82 n_3}{14} \left[\left(\frac{d}{R_1} \right)^2+\left(\frac{d}{R_2} \right)^2\right]\right.\right.
\nonumber \\
& \left.\left. \left.
-\frac{25+38 n_3}{7} \, \frac{d^2}{R_1 R_2}  \right\}\cos(2 \theta)+ \left(\frac{d}{R_2}-\frac{d}
{R_1}\right)\left[   6\,(1+2 n_3) -\frac{27+62 n_3}{7}\left(\frac{d}{R_1}+\frac{d}{R_2}\right)\right]\cos(2 \phi) \sin^2(\theta)\right\} \right\}\;.\label{potanggold}
\end{align}
\end{widetext}
This result shows again clearly that for a flat surface there is no orientation dependence of the potential. 
In Fig.~\ref{fig_stability_Au_pancake} we show the stable orientations of a prolate spheroid with $n_3=0.7$ (``pancake''), as a function of $d/R_1$ and $R_1/R_2$. In Fig.~\ref{fig_stability_Au_needle} an analogous plot is shown for a prolate spheorid with $n_3=0.2$ (``needle'').  It is interesting to note that there exist values of $R_1/R_2$ for which the stable orientation changes as the distance $d$ is varied. For a spherical surface ($R_1=R_2$) a ``pancake'' prefers to sit parallel to the surface (symmetry axis oriented along the $z$ axis) if it is located inside the sphere (negative radii of curvature) while a ``needle'' points towards a spherical surface if is outside the surface (positive radii of curvature). 
 
\begin{figure}[th]
\includegraphics [width=.9\columnwidth]{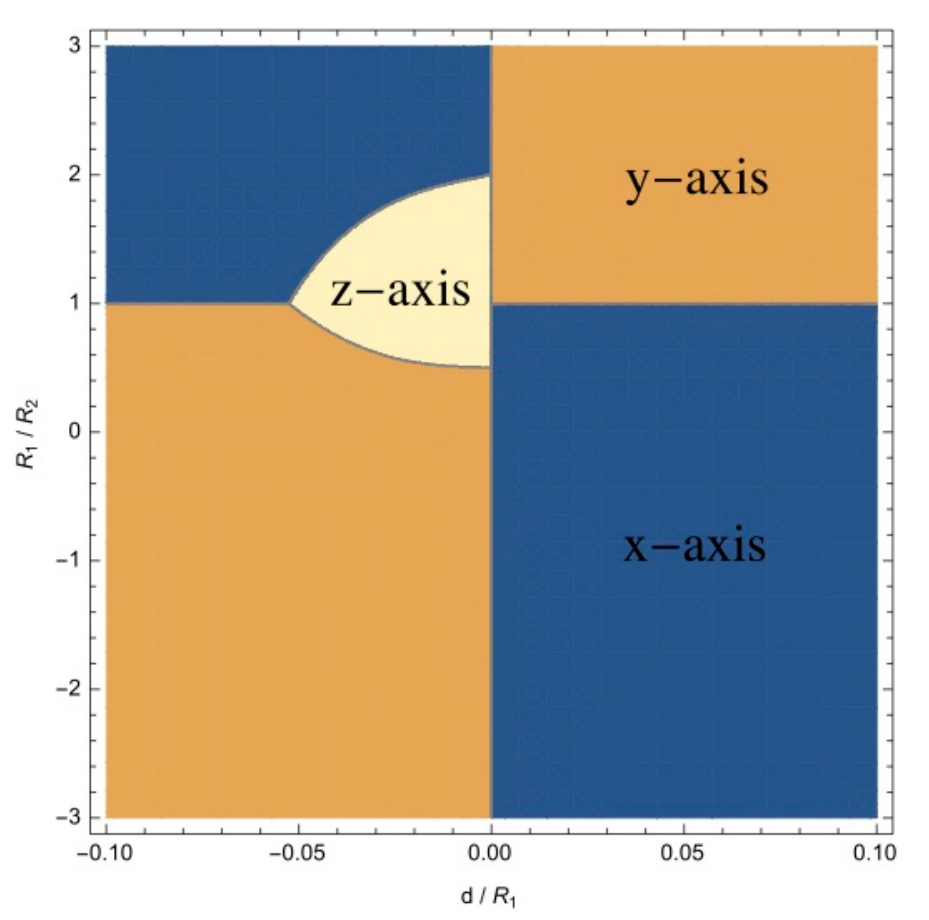} 
\caption{\label{fig_stability_Au_pancake_finiteT} Stability diagram for a gold  oblate spheroid (``pancake'') with $n_3=0.7$ and $R_1=20 \, \mu$m at $T=300$ K.}
\end{figure}
\begin{figure}[th]
\includegraphics [width=.9\columnwidth]{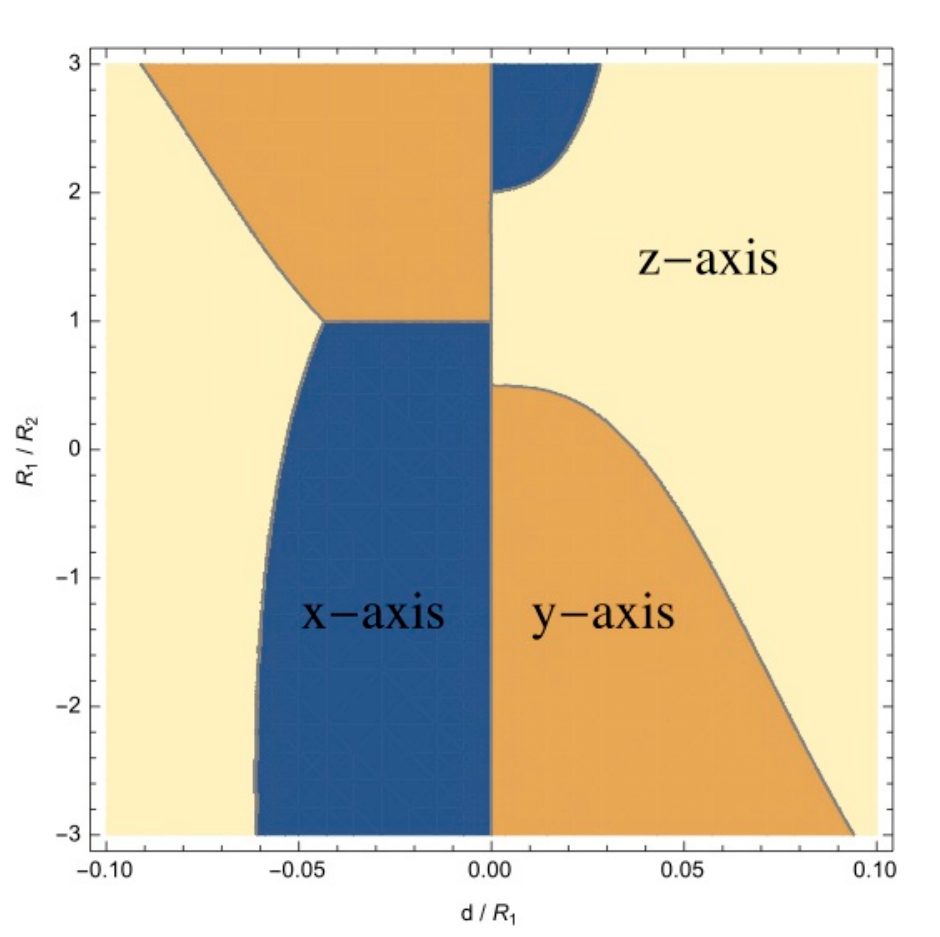} 
\caption{\label{fig_stability_Au_needle_finiteT} Stability diagram for a gold prolate spheroid (``needle'') with $n_3=0.2$ and $R_1=20\,\mu$m at $T=300$ K.}
\end{figure}

Finite temperatures  modify the stability diagrams: We consider again room temperature, $T=300$ K, and assume that $R_1=20 \, \mu$m (which we have to specify here explicitly since it is compared to the thermal wave length, contrary to the $T=0$ case). As can be observed from Figs.~\ref{fig_stability_Au_pancake_finiteT},~\ref{fig_stability_Au_needle_finiteT}, thermal fluctuations reduce the stability region for orientations of a ``pancake'' along the $z$-axis while  increasing the stability  for $z$-axis orientations of a ``needle.''
In the latter case there is a change of the preferred orientation for almost all ratios $R_1/R_2$ with increasing distance $d$ from either $x$- or $y$-orientation to a $z$-orientation.

\section{Conclusions \& Outlook}
On symmetry grounds it is expected that the Casimir--Polder force on an anisotropic 
particle, characterized by electric and magnetic dipolar polarizability tensors, 
should depend on its orientation relative to a nearby surface, with
a torque rotating the object to energetically favorable alignment.
Actually, for perfect conductors at zero temperature, and asymptotically at large distances,
the interaction depends only on the trace of the static polarizability tensor,
and orientation dependence at large separations is generically weak.
At short distances, comparable to the size of the object, strong orientation dependence
is inevitable, selecting a favorable alignment for contact.
For example, a prolate spheroid (pancake) will position itself with symmetry axis 
perpendicular to a flat surface ($z$ direction), 
while an oblate (cigar) one will have its axis parallel to the surface ($(x,y)$ plane).
A curved surface, with distinct radii of curvature, will then break the rotational degeneracy 
of the oblate spheroid parallel to the plate.

In this paper, we have studied the effects of surface curvature for the
Casimir--Polder force on anisotropic nano-particles.
The gradient expansion holds in an intermediate range of separations, 
larger than the particle size, but smaller than the radii of curvature.
While the expressions we find are quite generally valid-- for arbitrary polarizability
tensors and general material properties-- we have focused on the easily visualizable
case of spheroids near gently curved perfect conductors.
We find that the interplay of surface curvature and particle anisotropy leads
to an orientation dependent interaction which is quite sensitive to temperature,
separation, and dielectric response. 
While the minimum energy orientation is either perpendicular to the surface, or aligned
to one of principal axes of curvature, the preferred alignment can change with
temperature or separation to the surface.

It should be noted that the computed orientation--dependence is a small fraction of
the net Casimir--Polder interaction, complicating potential experimental probes:
Freely suspended particles will be absorbed by the substrate, while
trapped particles need to be cooled to very low temperatures before orientation
preferences can be manifested.
Nevertheless, it has been suggested~\cite{Thiyam15} that such forces may be implicated
in absorption properties of anisotropic molecules.
A quantum treatment of the problem, applicable to the scales of molecular adsorption, 
would thus be a valuable extension.




\begin{acknowledgments}
We thank R.~L.~Jaffe  for valuable discussions.  This
research was supported by the NSF through grant No. DMR-12-06323.
\end{acknowledgments}

\end{document}